\documentclass[twocolumn]{pasj00}

\begin{document}
\SetRunningHead{K.\ Wajima \etal}{Radiation Testing of COTS High-Speed
LSI Chips}
\Received{2006/12/15}
\Accepted{2007/08/24}

\title{Radiation Testing of Consumer High-Speed LSI Chips\\
for the Next Space VLBI Mission, VSOP-2}


\author{%
Kiyoaki \textsc{Wajima}\altaffilmark{1,2}
Noriyuki \textsc{Kawaguchi}\altaffilmark{3,4},
Yasuhiro \textsc{Murata}\altaffilmark{5,6},
and
Hisashi \textsc{Hirabayashi}\altaffilmark{5}}

\altaffiltext{1}{Korea Astronomy and Space Science Institute, 61-1 Hwaam-dong,
Yuseong, Daejeon 305-348, Korea}

\altaffiltext{2}{Department of Physics, Faculty of Science, Yamaguchi
University, 1677-1 Yoshida, Yamaguchi, Yamaguchi 753-8512}
\email{wajima@yamaguchi-u.ac.jp}

\altaffiltext{3}{National Astronomical Observatory of Japan, 2-21-1 Osawa,
Mitaka, Tokyo 181-8588}

\altaffiltext{4}{Department of Astronomical Science, The Graduate University
for Advanced Studies, \\ 2-21-1 Osawa, Mitaka, Tokyo 181-8588}

\altaffiltext{5}{Institute of Space and Astronautical Science, Japan Aerospace
Exploration Agency, \\ 3-1-1 Yoshinodai, Sagamihara, Kanagawa 229-8510}

\altaffiltext{6}{Department of Space and Astronautical Science, The Graduate
University for Advanced Studies, \\ 3-1-1 Yoshinodai, Sagamihara, Kanagawa
229-8510}


\KeyWords{space vehicles: instruments --- instrumentation: interferometers ---
techniques: interferometric}

\maketitle

\begin{abstract}

We performed two types of radiation testing on high-speed LSI chips to test
their suitability for use in wideband observations by the Japanese next
space VLBI mission, VSOP-2.
In the total ionization dose experiment we monitored autocorrelation spectra
which were taken with irradiated LSI chips and the source current at intervals
up to 1,000 hours from the ionization dose, but we could not see any change
of these features for the chips irradiated with dose rates expected in the
VSOP-2 mission.
In the single event effect experiment, we monitored the cross correlation
phase and power spectra between the data from radiated and non-radiated
devices, and the source current during the irradiation of heavy-ions.
We observed a few tens of single event upsets as discrete delay jumps for each
LSI.
We estimated the occurrence rate of single events in space as between once
a few days to once a month.
No single event latch-up was seen in any of the LSIs.
These results show that the tested LSIs have sufficient tolerance to the
environment for space VLBI observations.


\end{abstract}

\section{Introduction}
\label{sec:Introduction}

The VLBI (Very Long Baseline Interferometry) technique allows high-resolution
astronomical images to be obtained from observations using widely separated
radio telescopes \citep{Thompson01}.
This technique enables extremely high angular resolutions, of less than
1 milliarcsecond, to be obtained and intercontinental VLBI observations are
now routinely carried out \citep{Zensus97,Kellermann01}.

The technique of space VLBI uses an orbiting radio telescope(s) in addition
to ground radio telescopes to achieve even further improvements in
angular resolution.
The first space VLBI mission, VSOP (the VLBI Space Observatory Programme),
was realized with the space VLBI spacecraft HALCA (Highly Advanced Laboratory
for Communications and Astronomy), which was launched in 1997
(\cite{Hirabayashi98}; \yearcite{Hirabayashi00}).

Following the successes of VSOP, the next space VLBI mission, VSOP-2, was
planned and selected as a scientific mission, ASTRO-G, in the Japan Aerospace
Exploration Agency (JAXA) with a proposed launch in 2012
\citep{Hirabayashi04,Murata05}.
In this paper we report the results of radiation testing of high-speed
commercial off-the-shelf (COTS) LSI chips as one of development items for
higher sensitivity space VLBI observations.

The minimum sensitivity, $S_{\rm min}$, in VLBI observations of continuum
sources can be derived as
\begin{equation}
S_{\rm min} = 2 k_{\rm B} R_{\rm SN} \sqrt{\frac{T_{\rm sys1} T_{\rm sys2}}
{A_{\rm e1} A_{\rm e2}}} \frac{1}{\sqrt{2 \Delta B \tau}},
\label{eqn:Sensitivity}
\end{equation}
where $k_{\rm B}$ is Boltzmann's constant, $R_{\rm SN}$ is the signal-to-noise
ratio of the observation signal, $T_{{\rm sys}n}$, $A_{{\rm e}n}$ are the
system noise temperature and the antenna effective aperture in station $n$,
respectively, $\Delta B$ is the bandwidth, and $\tau$ is the integration time.
In VLBI, the radio signals are usually digitized and accumulated at the
Nyquist rate of $2\Delta B$, and thus one can detect fainter objects with
higher speed sampling.
Expansion of the bandwidth by raising the sampling speed is therefore a key
issue for high sensitivity VLBI observations.
This is particularly true for space VLBI, where the dimensions of the rocket
nose fairing place strong constraints on the antenna size.
The VSOP-2 spacecraft will have the capability of a data processing rate of
1\,gigabit per second (Gbps), as compared to 128\,Mbps for HALCA spacecraft.
While data sampling and recording at more than, or equal to, 1\,Gbps have
already been accomplished with ground-based VLBI systems \citep{Nakajima01},
it has not yet been realized in an onboard system, in part because to date
there have been very few missions that require such a high speed sampling
system.
Radiation testing of COTS LSI chips is therefore indispensable to realize
wideband space VLBI observations.

In addition, we must consider the orbit of the space VLBI spacecraft.
In space VLBI observations, higher quality radio images can be obtained by
placing the spacecraft in an elliptical orbit.
For example, HALCA had an elliptical orbit with an apogee height of
21,400\,km and a perigee height of 560\,km, and the VSOP-2 spacecraft will
have a similar orbit to that of HALCA, with nominal apogee and perigee
height of 25,000\,km and 1,000\,km, respectively.
However, it is a very severe environment for the spacecraft because such an
orbit passes through the inner Van Allen belt twice per revolution.
In 2002, the Mission Demonstration Satellite-1 (MDS-1, later named TSUBASA)
was launched by the National Space Development Agency of Japan (NASDA) and
placed in an elliptical orbit (an apogee height of 35,696\,km and a perigee
height of 500\,km), which was similar to that of HALCA.
MDS-1 was equipped with several COTS products, such as solar cells,
semiconductor and memory devices, in order to verify their tolerance in the
space environment \citep{Shindou02}.
However, it was not equipped with the high speed LSI chips required in the
VSOP-2 mission, and therefore radiation testing of such chips is necessary.

For the above-mentioned reasons, we performed radiation testing of high
speed COTS devices.
In section~\ref{sec:Experiments Summary}, we summarize the experiments.
Overviews and results of the two types of experiments are given in
sections~\ref{sec:TID Experiment} and \ref{sec:SEE Experiment}.
In section~\ref{sec:Discussion} we discuss the suitability of the tested
devices for the space environment.


\section{Summary of the Experiments}
\label{sec:Experiments Summary}

\begin{figure}
\begin{center}
\FigureFile(80mm,50mm){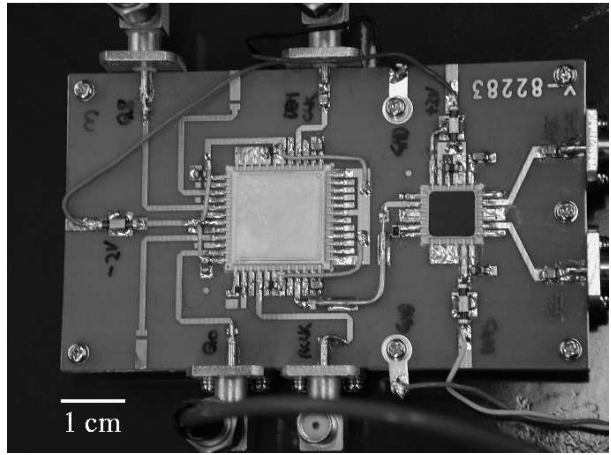}
\end{center}
\caption{Tested devices; a demultiplexer (left) and a decision circuit
(right).}
\label{fig:Figure1}
\end{figure}

The experiments were carried out at the Takasaki Institute of the Japan
Atomic Energy Research Institute (JAERI).
The tested devices were a decision circuit (1-bit quantization circuit;
hereafter DEC) and a demultiplexer (converting one data stream into
16-bit parallel data; hereafter DEMUX), both of which are gallium arsenide
(GaAs) MESFET devices having gate lengths of 0.2\,$\mu$m (see
figure~\ref{fig:Figure1}).
These are available for ground communications.
These have the capability of a maximum data processing rate of 10\,Gbps
and are used in an 8\,Gbps sampler which is under development at the National
Astronomical Observatory of Japan (NAOJ) \citep{Okiura02}.

We have performed two types of radiation testing; a total ionization
dose (TID) experiment using a cobalt-60 source, and a single event effect
(SEE) experiment by irradiation of heavy-ions.
If we assume a nominal orbit for the VSOP-2 spacecraft, as mentioned in
section \ref{sec:Introduction}, and nominally a one year lifetime, then
1\,kGy (an unit 'gray' represents an absorbed dose in the dimension of
J kg$^{-1}$) of total ionizing radiation and a maximum incidence level of
80\,MeV\,mg$^{-1}$\,cm$^2$ for the linear energy transfer (LET; energy loss
rate of a particle along its trajectory in a material) are required.


\section{Total Ionization Dose Experiment}
\label{sec:TID Experiment}

\subsection{Overview of the Experiment}
\label{subsec:TID Overview}

\begin{figure}
\begin{center}
\FigureFile(80mm,80mm){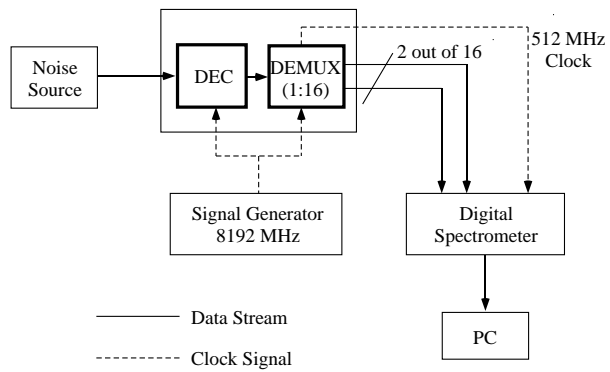}
\end{center}
\caption{Schematic diagram of the power spectra measurement system for the
TID experiment.}
\label{fig:Figure2}
\end{figure}

In the TID experiment six sets of DECs and DEMUXs were prepared, five sets
of which served as test devices and one of which was used as a reference.
Irradiation testing was carried out at a dose rate of 2\,kGy\,hr$^{-1}$,
and the absorbed dose toward each device measured by the dosimeter was
0.73\,kGy, 1.70\,kGy, 1.90\,kGy, 5.01\,kGy, and 9.95\,kGy.
After the irradiation, autocorrelation spectra were taken using the
measurement system shown in figure~\ref{fig:Figure2}.
In this system the white noise generated by the noise source is one-bit
digitized by the DEC and one data stream is parallelized into 16 streams
by the DEMUX.
In our experiment, however, only two data streams out of 16 parallel outputs
from the DEMUX were used for correlation due to limitations of the processing
speed after the DEMUX.

Autocorrelation spectra were taken seven times after the irradiation
(8, 24, 72, 100, 150, 300, and 1,000 hours) and compared with the spectrum
taken before irradiation.
The source current of each LSI was also monitored up to 1,000 hours from
the experiment, with  sampling times of 10 seconds during the irradiation
and an hour for monitoring after the experiment.


\subsection{Results}
\label{subsec:TID Results}

\begin{figure}
\begin{center}
\FigureFile(80mm,50mm){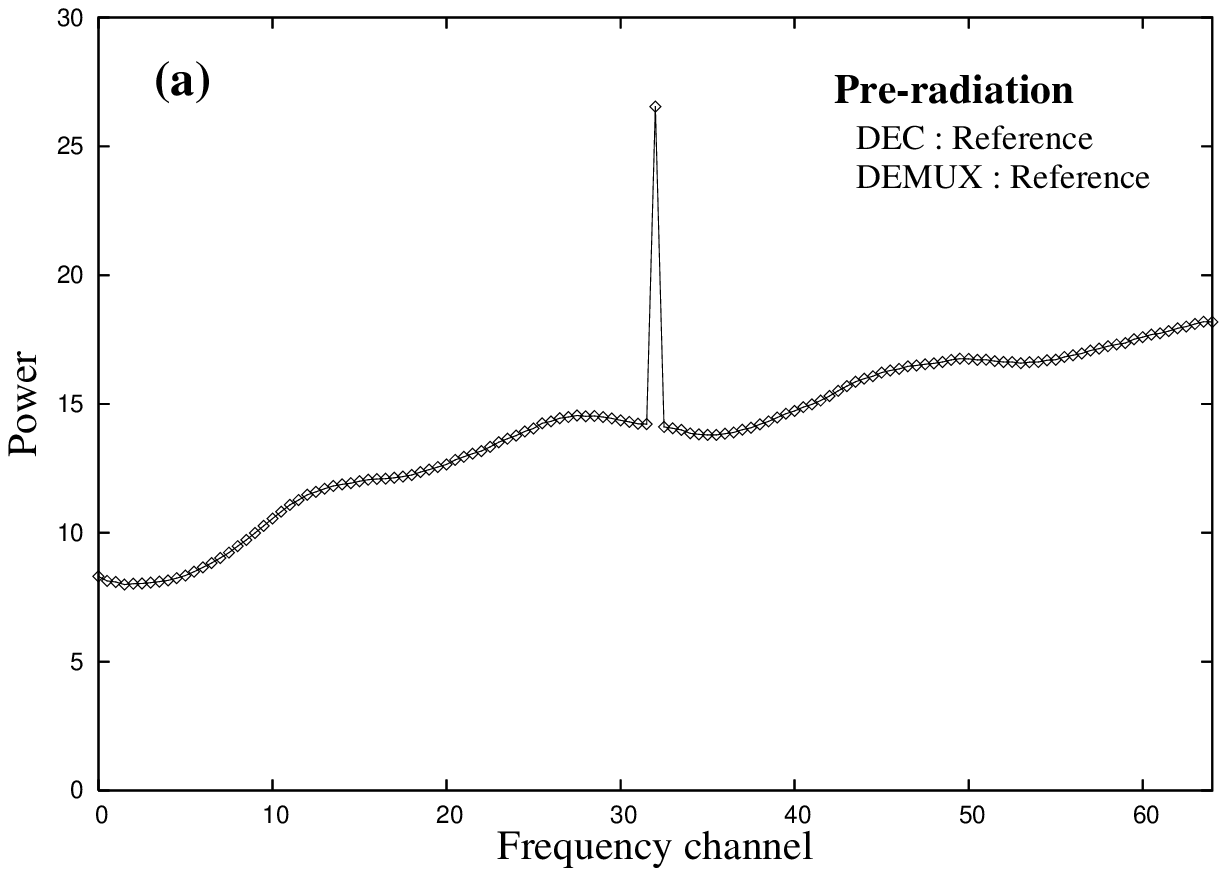}
\FigureFile(80mm,50mm){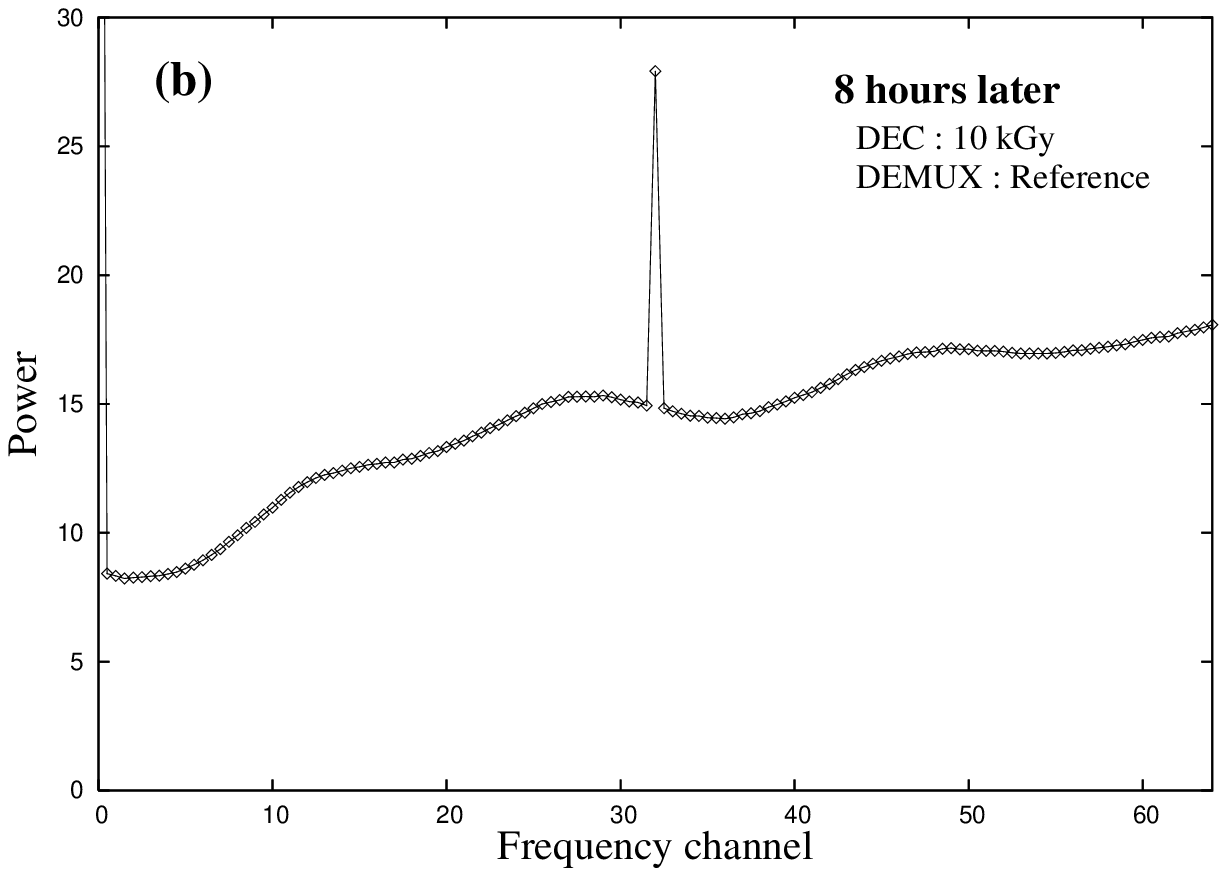}
\FigureFile(80mm,50mm){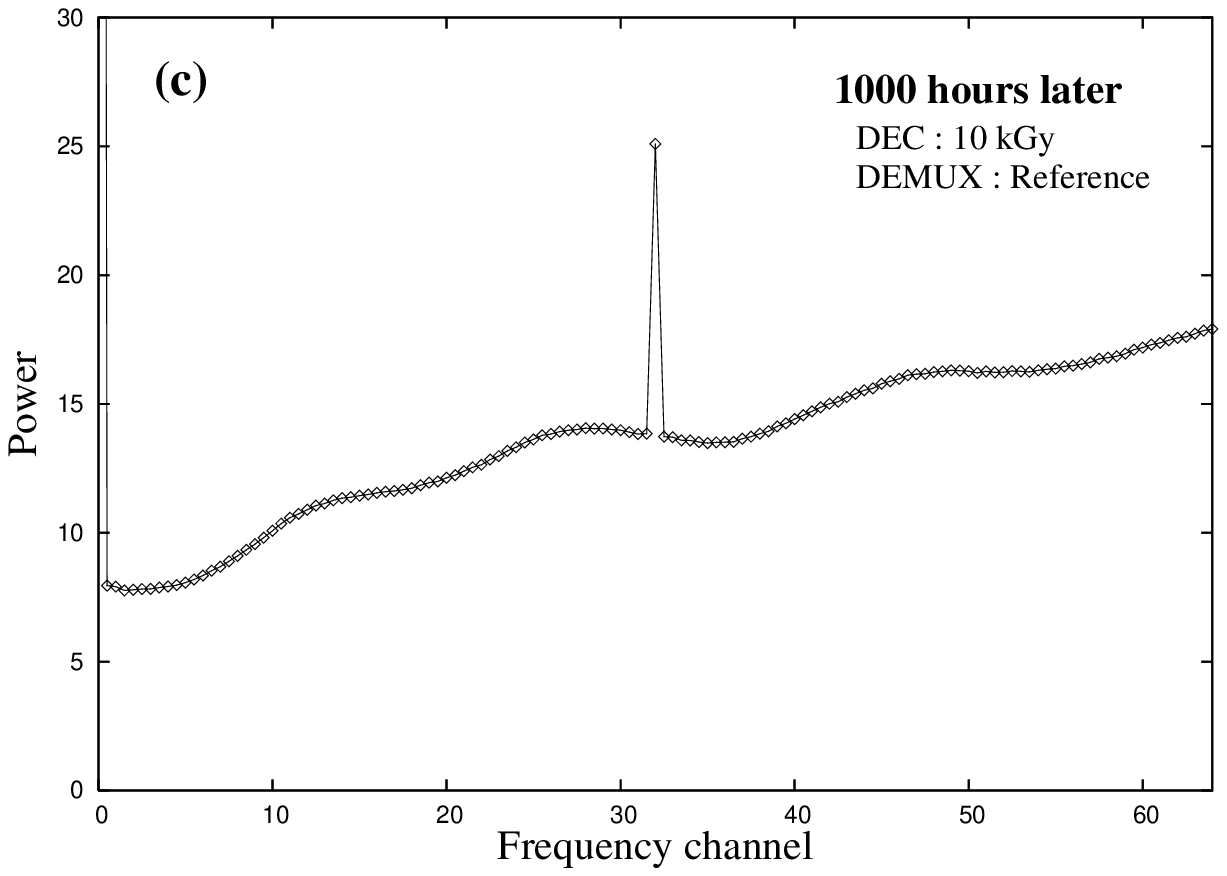}
\FigureFile(80mm,50mm){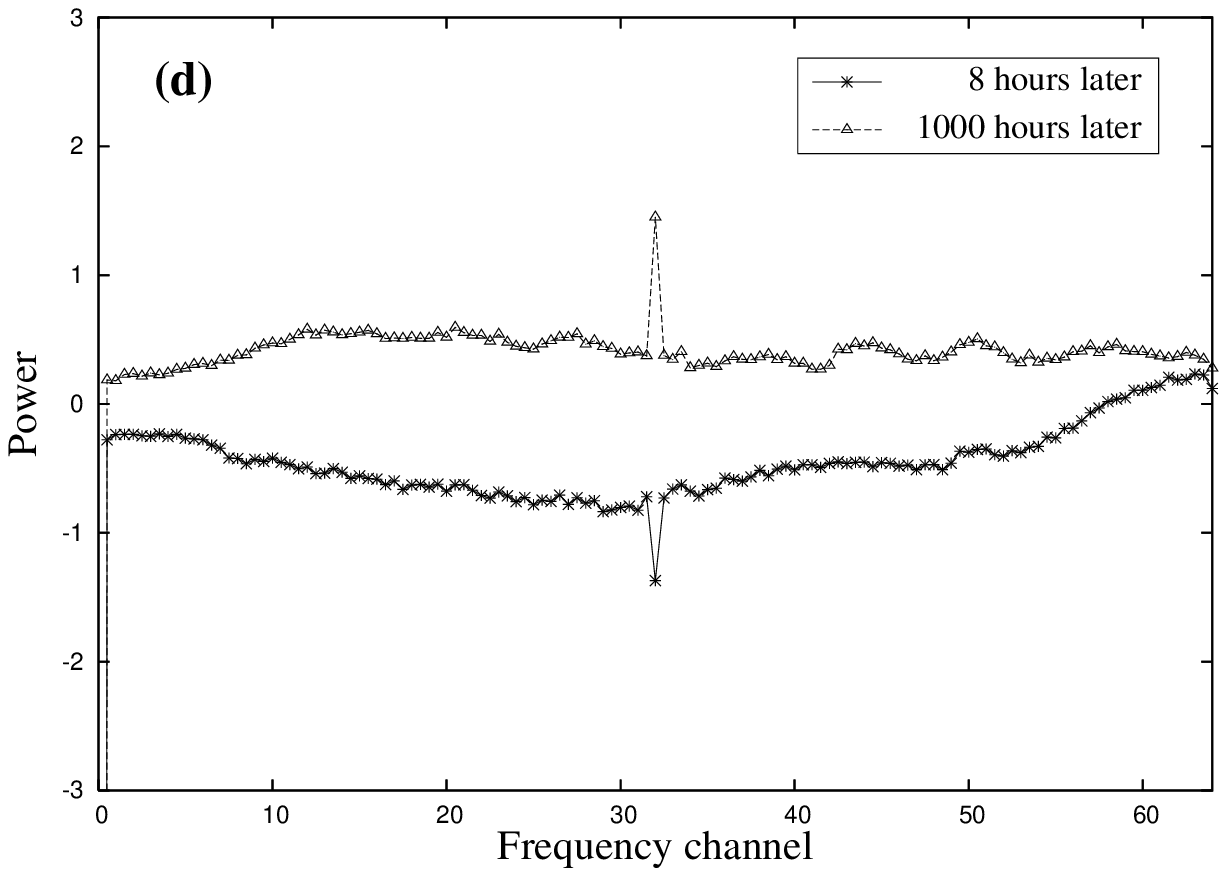}
\end{center}
\caption{Power spectra in the TID experiment. (a) Pre-radiation,
(b) 8 hours after the experiment; DEC: 10 kGy irradiation, DEMUX:
reference, (c) 1,000 hours after the experiment; DEC: 10 kGy irradiation,
DEMUX: reference,
(d) result of subtraction of panel (b) and (c) from panel (a),
used as a reference.}
\label{fig:Figure3}
\end{figure}

Part of autocorrelation spectra results from the TID experiment are shown
in figure~\ref{fig:Figure3} (a), (b), and (c).
The horizontal and vertical axes show the frequency channel (one channel
corresponds to a bandwidth of 4\,MHz) and the power in arbitrary units,
respectively.
The peak at channel 32 in each graph is a monochromatic signal injected as
a reference and it is adjusted to a similar power level to the wideband noise.
Figure~\ref{fig:Figure3} (d) shows the result of subtraction of power in
panel (b) and (c) from panel (a), assuming as a reference.
The standard deviation for each graph compared to (a), excluding channels
0 (DC offset can be seen) and 32 (a monochromatic signal is injected) is
0.260 and 0.095, showing little change in amplitude before and after
irradiation.
In order to check the influence of irradiation, we also calculated the
cross-correlation coefficient between the power spectra of the reference and
the tested device.
In all combinations, including the case illustrated by
figure~\ref{fig:Figure3}, cross-correlation coefficients are larger than
0.993, corresponding to the amplitude displacement of less than 0.03\,dB.
If this amplitude displacement is produced by the TID effect over the
lifetime of VSOP-2 mission, we conclude that the displacement within a
typical duration time per observation (about 8 hours) can be ignored.

\begin{figure}
\begin{center}
\FigureFile(80mm,50mm){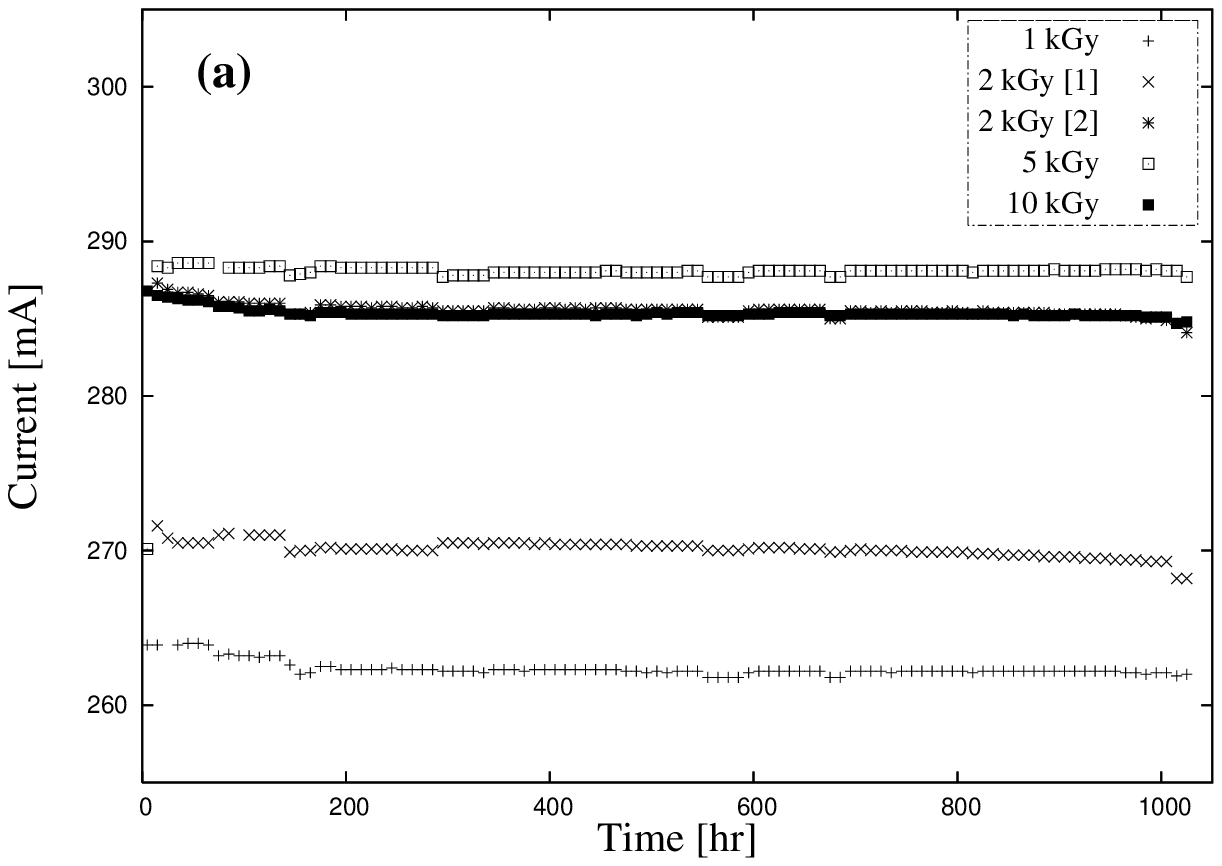}
\FigureFile(80mm,50mm){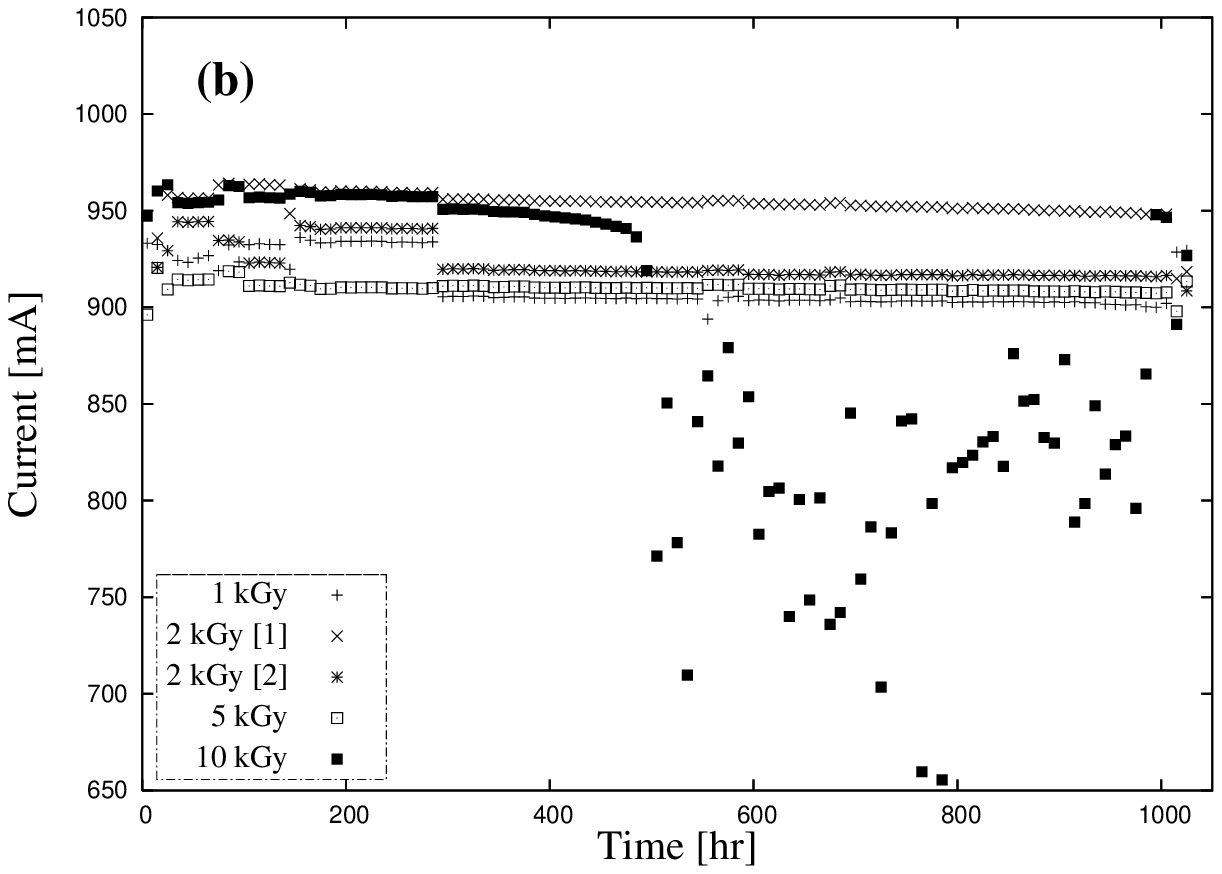}
\end{center}
\caption{The source current in the TID experiment; (a) DEC and (b) DEMUX.}
\label{fig:Figure4}
\end{figure}

Figure~\ref{fig:Figure4} shows the source current of all DEC and DEMUX
samples during and after the experiment.
There are several discontinuities in each of the sampled data because it was
necessary to mount and unmount  each LSI in order to take autocorrelation
spectra.
In all DECs we could see slight variations in the source current up to 1\%.
On the other hand, an abrupt current variation in the DEMUX with 10\,kGy
irradiation was observed 500 hours after the experiment.
We could not determine whether this variation was due to the excess of
irradiation, or to gamma-ray irradiation itself irrespective of the total
dose.
We should perform a further experiment in order to specify the cause,
although, if the former case applies, we can conclude that these LSIs
are able to be used in space as the total dose of 10\,kGy corresponds to
that 10 times longer than the planned lifetime of the VSOP-2 spacecraft.



\section{Single Event Effect Experiment}
\label{sec:SEE Experiment}

\subsection{Overview of the Experiment}
\label{subsec:SEE Overview}

\begin{figure}
\begin{center}
\FigureFile(80mm,80mm){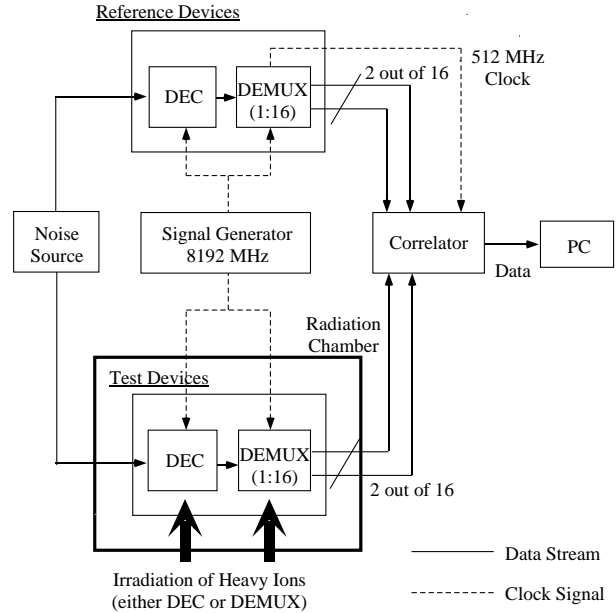}
\end{center}
\caption{Schematic diagram of the cross correlation measurement system for the
SEE experiment.}
\label{fig:Figure5}
\end{figure}

In the SEE experiment, four sets of DECs and DEMUXs were prepared, three sets
of which served as tested devices and one set of which was the reference.
Heavy ions, krypton (Kr, ${\rm LET} = 6.33$\,MeV\,mg$^{-1}$\,cm$^2$), argon
(Ar, ${\rm LET} = 15.3$\,MeV\,mg$^{-1}$\,cm$^2$), and neon
(Ne, ${\rm LET} = 39.9$\,MeV\,mg$^{-1}$\,cm$^2$), were used to irradiate
either DEC or DEMUX.
Because of a problem in the beam line system, experiments using higher LET
ions,  such as xenon (Xe, ${\rm LET} = 62.9$\,MeV\,mg$^{-1}$\,cm$^2$), were
not performed.
In order to check the influence of the heavy-ion irradiation, the
cross-correlation between the data from radiated and non-radiated devices
were monitored in real-time during the experiment using the measurement
system shown in figure~\ref{fig:Figure5}.
As in the TID experiment, DECs and DEMUXs were driven with an 8\,GHz clock
and two data streams out of 16 parallel outputs from DEMUX were used for
cross-correlation.
The source current of the devices was also monitored during the irradiation
in order to check for the occurrence of a single event latch-up.

In the standard radiation testing of memory devices, known data are
stored into a memory before irradiation and the number of inverted bits is
counted after the testing.
One therefore cannot check the influence of the irradiation effect in
real-time, while in our measurement method one can directly monitor the
influence of the heavy-ion irradiation.

Before the experiment we checked the cross correlation spectrum under the
non-radiated condition and confirmed that the cross-correlation phase had
no variation with time.


\subsection{Results}
\label{subsec:SEE Results}

\begin{figure}
\begin{center}
\FigureFile(80mm,50mm){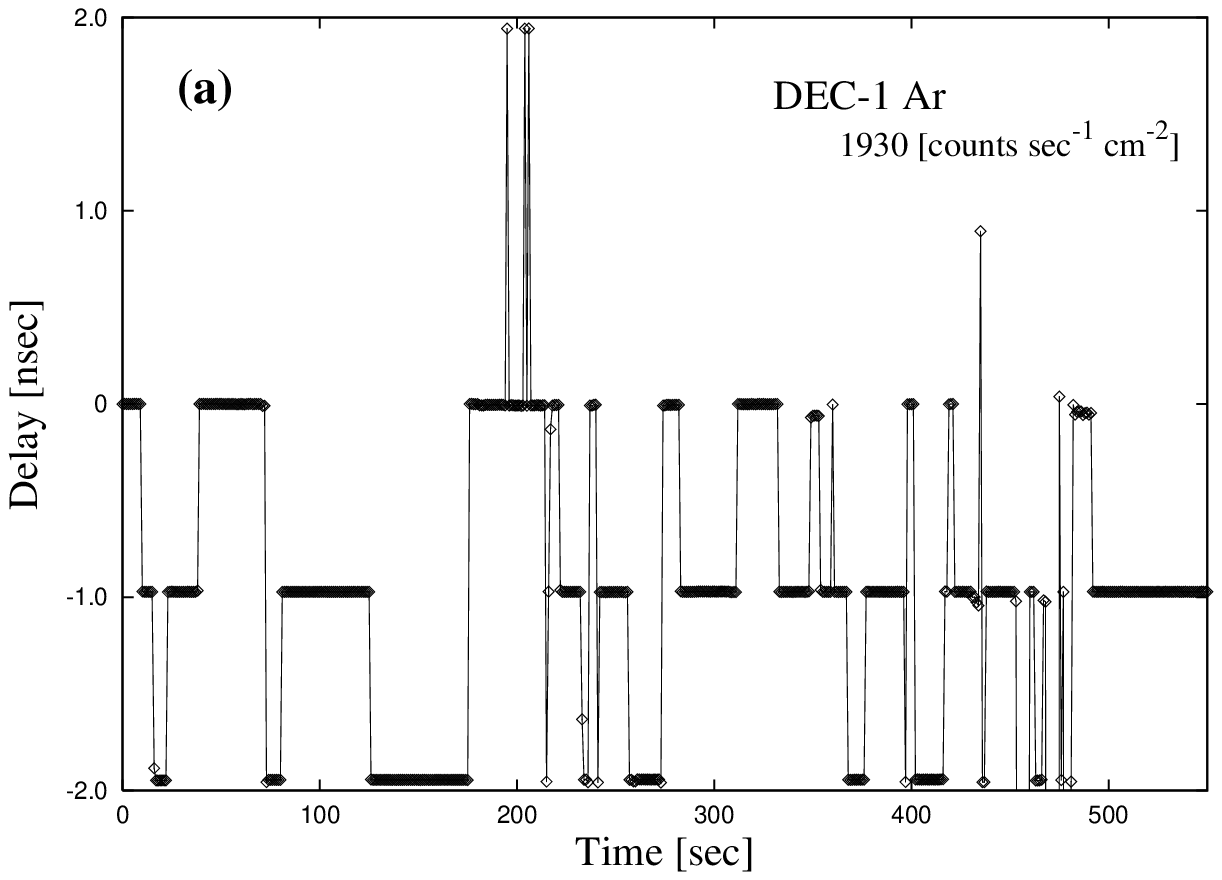}
\FigureFile(80mm,50mm){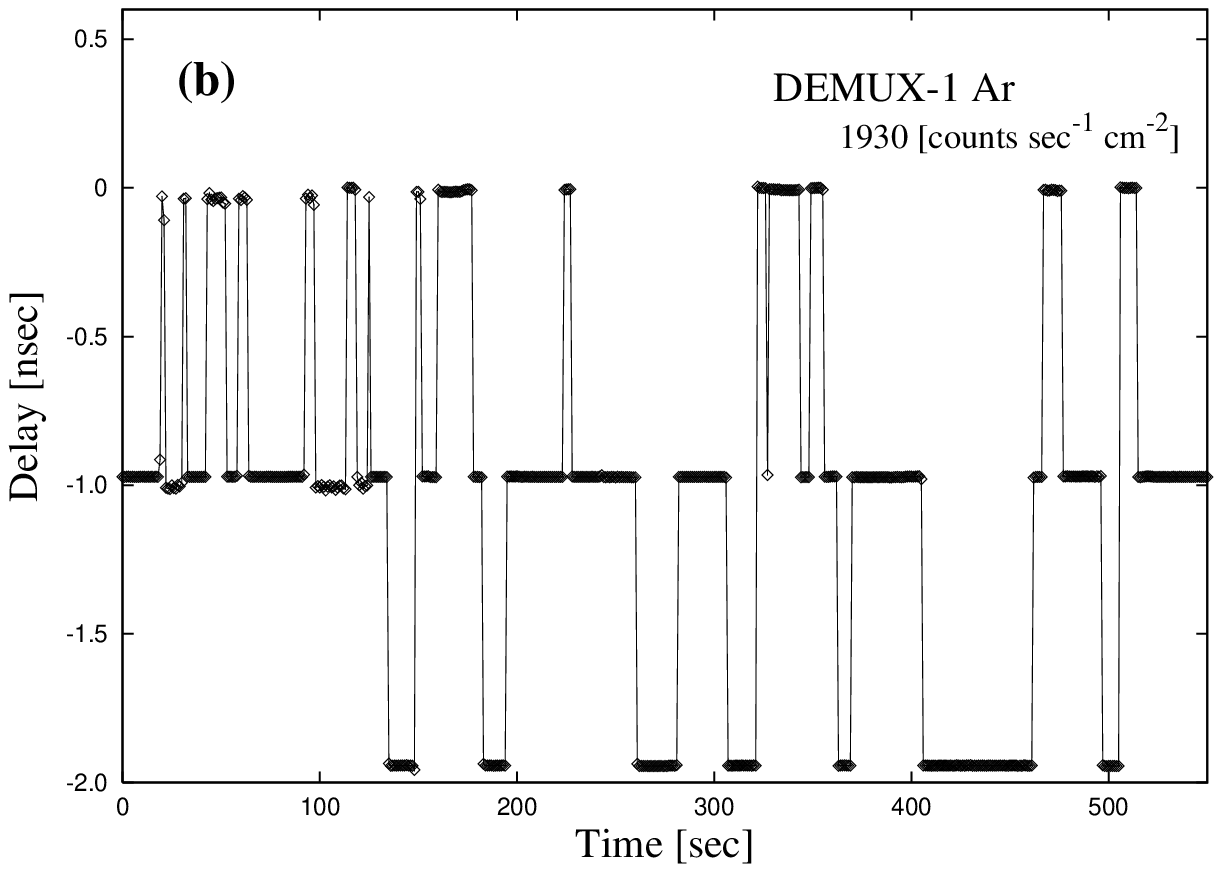}
\end{center}
\caption{Measured delay for the Ar irradiation of (a) DEC and (b) DEMUX.}
\label{fig:Figure6}
\end{figure}

Part of the result in the SEE experiment is shown in
figure~\ref{fig:Figure6}.
The horizontal and vertical axes show the radiation time in seconds and
the delay time between the reference and tested devices in nanoseconds,
respectively.
In the experiment shown in figure~\ref{fig:Figure6}, Ar was irradiated
at a rate of 1,930 particles\,cm$^{-2}$\,sec$^{-1}$, for a total of
$10^6$ particles.
The delay time was derived from the observation result of the time variation
of the cross-correlation phase, and a moving image of the cross-correlation
phase in the real-time observation result is available online
\footnote{See http://wwwj.vsop.isas.jaxa.jp/vsop2/radtest/Radtest.html}.

In the SEE experiment we had 12 -- 53 single event upsets (SEUs) for each LSI
and heavy-ion (see table~\ref{tbl:Table1}).
\begin{table}
\begin{center}
\caption{SEU counts for each heavy ion.$^*$}
\label{tbl:Table1}
\begin{tabular}{cccc}
\hline\hline
                                                     & Ne & Ar & Kr \\
\cline{2-4}
Particle injection rate                              &
\raisebox{-1.2ex}[0pt]{5,153} &
\raisebox{-1.2ex}[0pt]{1,930} &
\raisebox{-1.2ex}[0pt]{783} \\
\raisebox{0.6ex}[0pt]{[counts cm$^{-2}$ sec$^{-1}$]} &    &    &    \\
Total number of particles                            &
\raisebox{-1.2ex}[0pt]{10}    &
\raisebox{-1.2ex}[0pt]{10}    &
\raisebox{-1.2ex}[0pt]{2.3} \\
\raisebox{0.6ex}[0pt]{[$\times 10^5$ counts cm$^{-2}$]} &    &    &    \\ \hline
DEC-1                                                & 12 & 32 & 16 \\
DEC-2                                                & -- & 24 & -- \\
DEC-3                                                & -- & 29 & -- \\
DEMUX-1                                              & 12 & 35 & 20 \\
DEMUX-2                                              & -- & 38 & -- \\
DEMUX-3                                              & -- & 53 & -- \\
\hline
\multicolumn{4}{@{}l@{}}{\hbox to 0pt{\parbox{80mm}{\footnotesize
\par\noindent
$^*$ -- denotes experiments not executed.
}\hss}}
\end{tabular}
\end{center}
\end{table}
Table \ref{tbl:Table1} shows that the SEU counts of the DEMUX are slightly
larger than, or equal to, those of the DEC in all combinations of ions.
This result may reflect differences in tolerance for the space environment
between the DEC and DEMUX, although this conclusion is necessarily somewhat
tentative because of limited number of samples.
We also found SEUs as discrete $n\pi$ jumps in the cross-correlation phase.

\begin{figure}
\begin{center}
\FigureFile(80mm,50mm){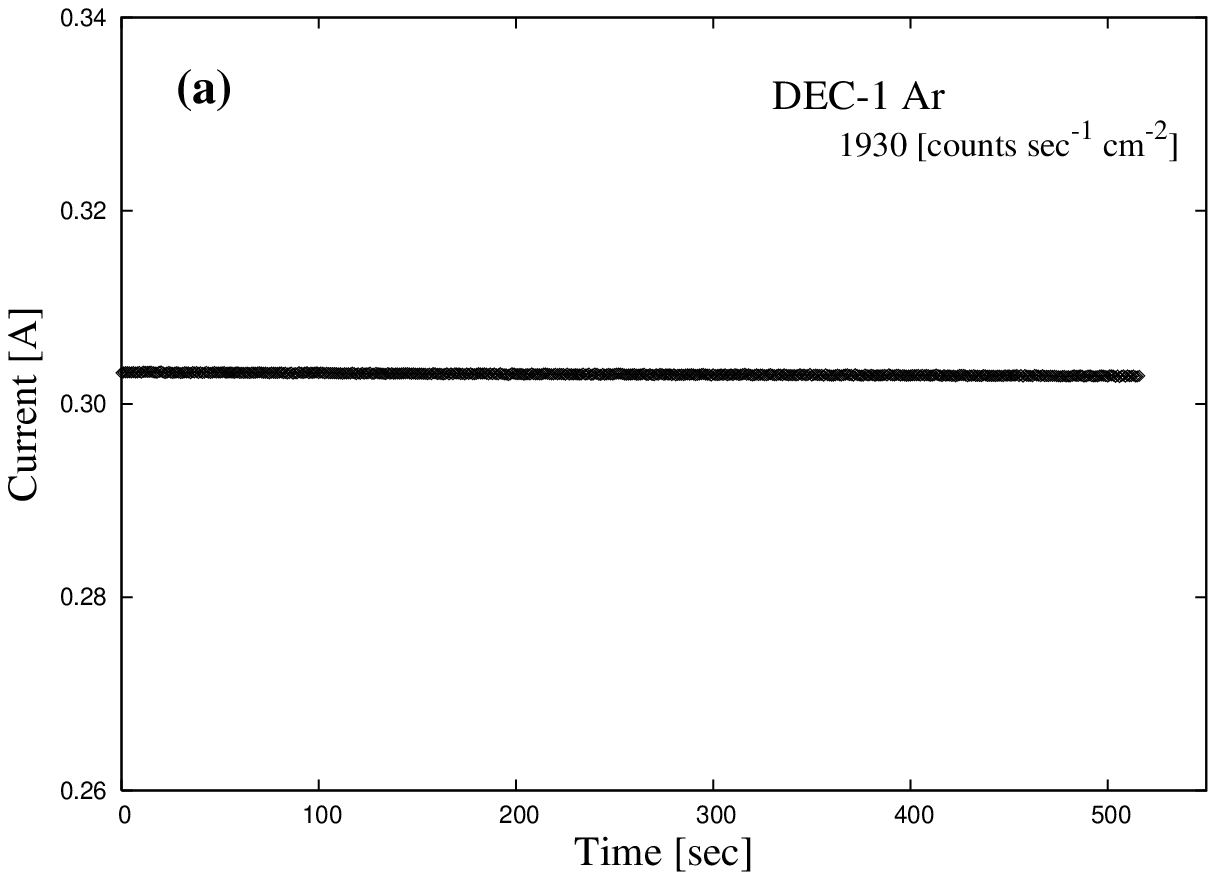}
\FigureFile(80mm,50mm){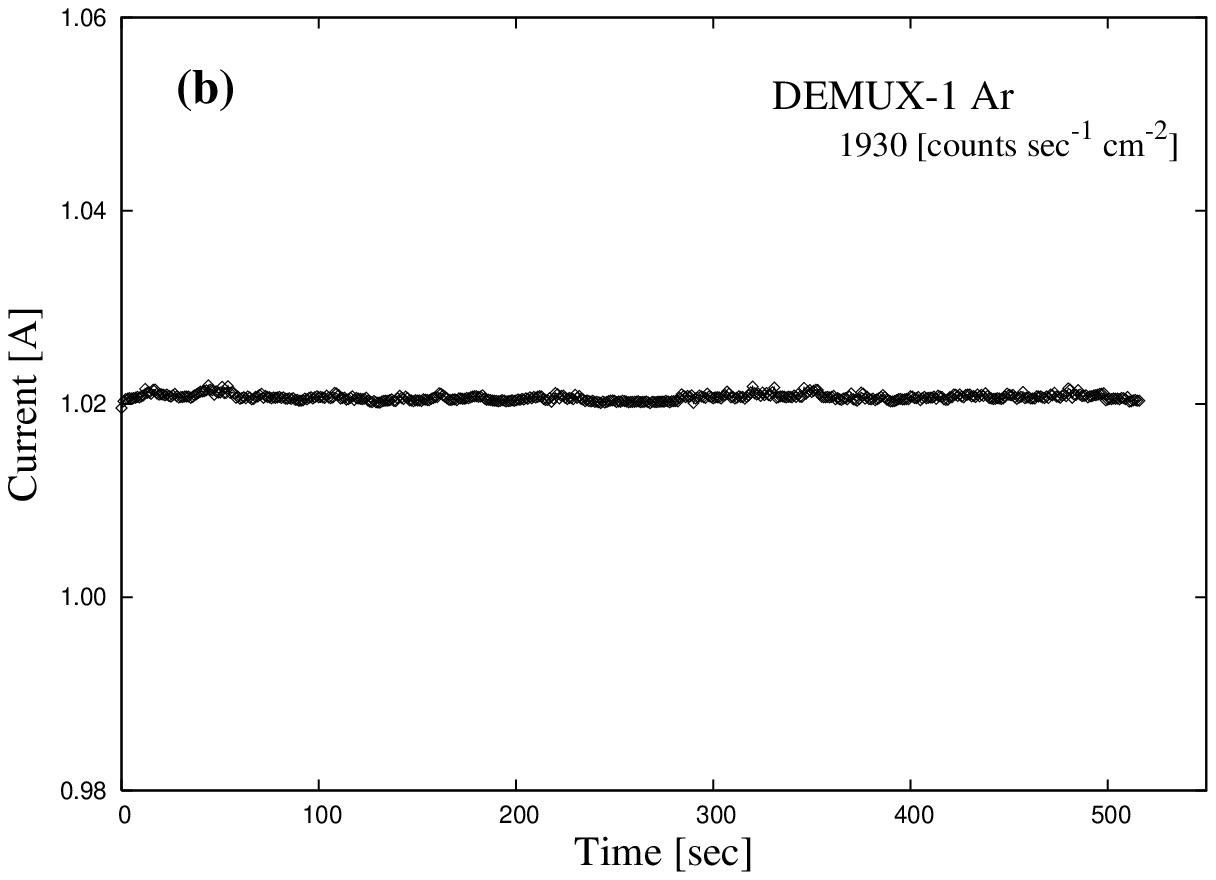}
\end{center}
\caption{Source current during Ar irradiation; (a) DEC and (b) DEMUX.}
\label{fig:Figure7}
\end{figure}

We have also monitored the source current of each LSI under radiation,
but did not see an abrupt change for any of LSIs
(see figure~\ref{fig:Figure7}).



\section{Discussion}
\label{sec:Discussion}

As mentioned in section~\ref{sec:SEE Experiment}, we could see SEUs as
discrete $n\pi$ jumps of the cross-correlation phase in both the DEC and
DEMUX data.
In the SEE experiment the DEC and DEMUX chips of both the reference and
tested paths were
driven by the 8\,GHz clock and two data streams, Q0 and Q8, out of 16
parallel outputs (Q0, Q1, $\cdot\cdot\cdot$, Q15 in order of time) from the
DEMUX were used for cross-correlation (see figure~\ref{fig:Figure5}).
Data processing was therefore done equivalently by the sampling rate of
1\,Gbps, and a displacement of one data sample between each data stream
corresponds to a one nanosecond delay.
From these observation results we consider that observed SEU in the DEMUX
is due to the occurrence of a clock reset in the DEMUX by the heavy-ion
injection.
On the other hand, we could also see discrete $n\pi$ jumps of the
cross-correlation phase in irradiation of the DECs, even though the clock
output was not used to drive the devices downstream (see
figure~\ref{fig:Figure5}).
We consider that the heavy-ion injection affects an internal register
circuit of the DEC giving rise to an SEU.

\begin{figure}
\begin{center}
\FigureFile(80mm,50mm){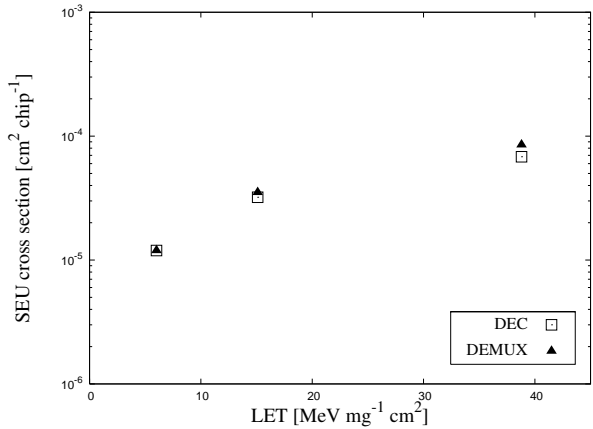}
\end{center}
\caption{SEU cross section of DEC-1 and DEMUX-1 for each heavy-ion.}
\label{fig:Figure8}
\end{figure}

In the SEE experiment we had a few tens of SEUs for each LSI and heavy-ion.
We derived the following `SEU cross-section' for each heavy-ion;
\begin{eqnarray*}
\lefteqn{{\rm SEU \,\, cross \,\, section \,\, [cm^2 \,\, chip^{-1}]}} & & \\
& = &\frac{{\rm total \,\, SEU \,\, number \,\, [counts\,\,chip^{-1}]}}
{{\rm number \,\, of \,\, the \,\, heavy \,\, ion \,\, particles \,\,
[counts\,\,cm^{-2}]}}.
\end{eqnarray*}
Figure~\ref{fig:Figure8} shows the relation between the LET and the SEU
cross-section for DEC-1 and DEMUX-1 for each heavy-ion.

We estimate the occurrence rate of SEUs for each LSI in space using the
CREME96 (Cosmic Ray Effects on Micro-Electronics) Program
\citep{Tylka97}\footnote{See also https://creme96.nrl.navy.mil/} with the
following procedure.
First, the critical charge $Q_{\rm c}$ [pC], the minimum charge deposition
required to produce an SEU, is calculated by the following:
\begin{equation}
Q_{\rm c} = \frac{L_{\rm th} T \rho e}{E},
\label{eqn:Critical Charge}
\end{equation}
where $L_{\rm th}$ is the threshold effective LET in MeV\,mg$^{-1}$\,cm$^2$,
$T$ is the device thickness in $\mu$m, $\rho$ is the material density
(5.32\,g\,cm$^{-3}$ for GaAs), $e$ is the elementary electric charge, and
$E$ is the minimum energy needed to create one electron-hole pair
(4.8\,eV for GaAs).
We derived $Q_{\rm c}$ from the device size and the expected $L_{\rm th}$,
which is equal to one hundredth of the saturated SEU cross section,
$\sigma_{\rm s}$, from figure~\ref{fig:Figure8}.
Finally, we can calculate the occurrence rate of SEU using derived $Q_{\rm c}$
along with orbital elements of the VSOP-2 spacecraft.
Although we cannot precisely derive $L_{\rm th}$ because we have not
performed the radiation testing using heavy-ions with lower LETs, if we
assume $L_{\rm th}$ to be 1, 2, 3, or 4 (MeV\,mg$^{-1}$\,cm$^2$), the
occurrence rate of SEU for both DEC and DEMUX can be estimated as about
once per 3.5, 14.1, 31.5, and 56.1 days, respectively.

The occurrence rate of SEU for each LSI seems to be slightly higher than
that for other space-qualified devices, however we believe that it is
acceptable for space VLBI observations for the following reason.
In VLBI, the digitized radio-astronomical data are usually recorded on a
magnetic storage media at each station and cross-correlation processing is
performed by a correlator in order to detect interferometric fringes.
When performing the cross-correlation, the expected arrival time of
identical wavefront at each station is calculated on the basis of the
coordinates of the radio source, the position of each antenna, and earth
rotation parameters, and those are subtracted from the original data in
advance.
Finally we search for the maximum cross-correlation coefficient in the
residual delay ($\tau_{\rm res}$) and residual delay-rate
($\dot{\tau}_{\rm res}$) plane.
In space VLBI there are  relatively large errors in the station position of
the spacecraft compared to those of ground radio telescopes because of the
orbital motion, and so correlators used for processing of space VLBI
observations use wider  $\tau_{\rm res}$ and $\dot{\tau}_{\rm res}$ ranges
to raise the chance of fringe detection.
For example, the Space VLBI correlator in NAOJ has the capability of a
maximum $\tau_{\rm res}$ of 128\,$\mu$sec, and one can carry out time
integration if fringes are found within this $\tau_{\rm res}$ window.
A few nanoseconds delay, which was observed in the SEE experiment, is much
smaller than the $\tau_{\rm res}$ window of the correlator and therefore
we can perform space VLBI observations using these LSI chips.
Even if one nanosecond delay observed in the SEE experiment occurs through
observations, the coherence loss in VLBI observations of continuum sources
is about 3\% and it is acceptable to accomplish scientific goals by the
VSOP-2 mission.

We thus have obtained promising prospects for the use of these LSIs in the
VSOP-2 project through these experiments, however we have not performed
the SEE experiment on the condition of higher LET irradiation with Xe or
of lower LET irradiation with protons or oxygen.
These are future tasks in order to determine $L_{\rm th}$ and $\sigma_{\rm s}$
precisely.


\section{Conclusion}
\label{sec:Conclusion}

We performed two types of radiation testing of high-speed LSI chips for
wideband space VLBI observations.
In these experiments we monitored the autocorrelation (TID) and
cross-correlation (SEE) spectra and the source current.
In the SEE experiment a few tens of single event upsets were seen for each
LSI, and we estimated the occurrence rate of single events in space as
between once a few days to once a month, which does not affect space
VLBI observations.
We did not see an abrupt change of the source current in any of LSIs.

The measurement method applied in the SEE experiment is useful for
radiation testing of similar types of LSI chips since one can monitor
the appearance of SEUs in real-time.

\bigskip

The authors would like to thank Drs.\ Sachiko K.\ Okumura and Satoru Iguchi
of NAOJ, Japan Communication Equipment (Nihon Tsushinki) Co.\ Ltd., and
High-Reliability Components (HIREC) Co.\ Ltd.\ for valuable comments and
technical supports on the experiments.
We are grateful to the referee, Dr.\ Jonathan D.\ Romney, for valuable
comments to improve the paper.
We also thank Dr.\ Philip G.\ Edwards for polishing the manuscript.




\end{document}